# PV Source Integrated Micro-Grid for Power Quality Improvement using MPPT Technique

**1 author:**

Niteesh Shanbog
PES Institute of Technology

# PV Source Integrated Micro-Grid for Power Quality Improvement using MPPT Technique

Madhu Palati[1], Manjunath R.D[2], Nagesh L[2], Niteesh S[2],Prashanth C[2]
[1]*Assistant Professor, Department of Electrical and Electronics Engineering, BMS Institute of Technology & Management, Bangalore, India*
[2]*Final year Students, Department of Electrical and Electronics Engineering, BMS Institute of Technology & Management, Bangalore, India*

**Abstract:** The demand for Electrical energy is increasing day by day as it can be easily converted to another form of energy. All consumers expect Electrical energy with high power quality. Most of the commercial and industrial loads are inductive in nature and need power electronic circuits/ controllers to get smooth control of the equipment. This in turn leads to the injection of harmonics into the system, hence the power quality is affected. The above problem needs to be addressed and eliminated. In this paper, shunt active power filter is used to mitigate the harmonics. $I_d$-$I_q$ control is used to analyse the performance of the filter and is simulated using MATLAB software. MPPT controller is used to improve the power quality of the system.

## I. INTRODUCTION

The Electrical energy is the most efficient and popular form of energy available, which can be easily converted into any other form. As most of the industrial and commercial loads like Fluorescent lamps, electronic converters, arc furnaces, etc., are nonlinear which require power electronic devices for smooth conversion and control. These devices lead to nonlinearity, unbalance condition, increased losses, poor power factor, poor efficiency, transient and steady state disturbance. Voltage harmonics and power distribution problem arises due to the current harmonics generated by nonlinear loads. These currents flow through the distribution lines causing more power loss, electromagnetic interference in the communication lines, distortion in the voltage waveform [1]. Therefore, the above problems lead to poor power quality and ultimately end consumer gets affected.

To mitigate the above problems conventionally passive filters consisting of LC filters or high pass filters were used [2]. Because of limitations like bulk size, resonance problem and only a few harmonics are eliminated. Alternate solution to mitigate the problems of passive filters is by using active filters. These are widely used because of its simple construction, excellent performance characteristics in single phase as well as in three phase system [2-4]. Active filters compensate harmonics in addition to asymmetric currents caused by unbalanced and non-linear loads. Shunt configuration of this filter directly injects current to cancel the harmonic current. Series configuration compensates the voltage distortion caused by non-linear loads. Based on requirement combination of the shunt and Series can be used to improve the performance of the filter.

Control strategy for generating compensating signals in terms of voltage or current is based on frequency domain or time domain analysis. Frequency domain analysis involves more calculations, complicated and uses Fourier transformation . p-q theory is the instantaneous reactive power theory can be used for steady state, transient operation. Only Algebraic expressions are involved, hence the calculations becomes simple [5]. This theory is based on transformation of three phase voltages in a-b-c coordinates to $0 - \alpha - \beta$ system.

In Instantaneous Active and Reactive Current Component ($i_d$-$i_q$) method, reference currents are generated through the instantaneous active and the reactive current component of the nonlinear load. The three phase current component a-b-c transform into α-β-0 components. The transformation angle 'θ' is obtained directly from main voltage. Synchronization problem can be avoided during unbalanced and non-sinusoidal voltage condition. Therefore, $i_d$-$i_q$ achieves large frequency operating limit.

## II. DESIGN OF SHUNT ACTIVE POWER FILTER

As most of the generation happening in India is through thermal generation, which leads to main global warming. Therefore, there is a need for robust, sustainable and environmental friendly power generation. Solar Photovoltaic (SPV) is the alternate source of energy which is pollution free, sustainable, environmentally friendly and available abundant in nature. Hence there is a need to integrate SPV generation with Power grid. There are many challenges like harmonics, low power factor, voltage fluctuations,, demand for reactive power,load variations etc. to integrate SPV generation with grid. Due to requirement of large storage battery, and above challenges, most of the research has focused on the low power generation only [6].The configuration of SAPF with Grid and schematic diagram of SPV generation interconnected to the grid using three Phase Voltage source converter (VSC) is shown in fig.1 and fig.2 respectively.

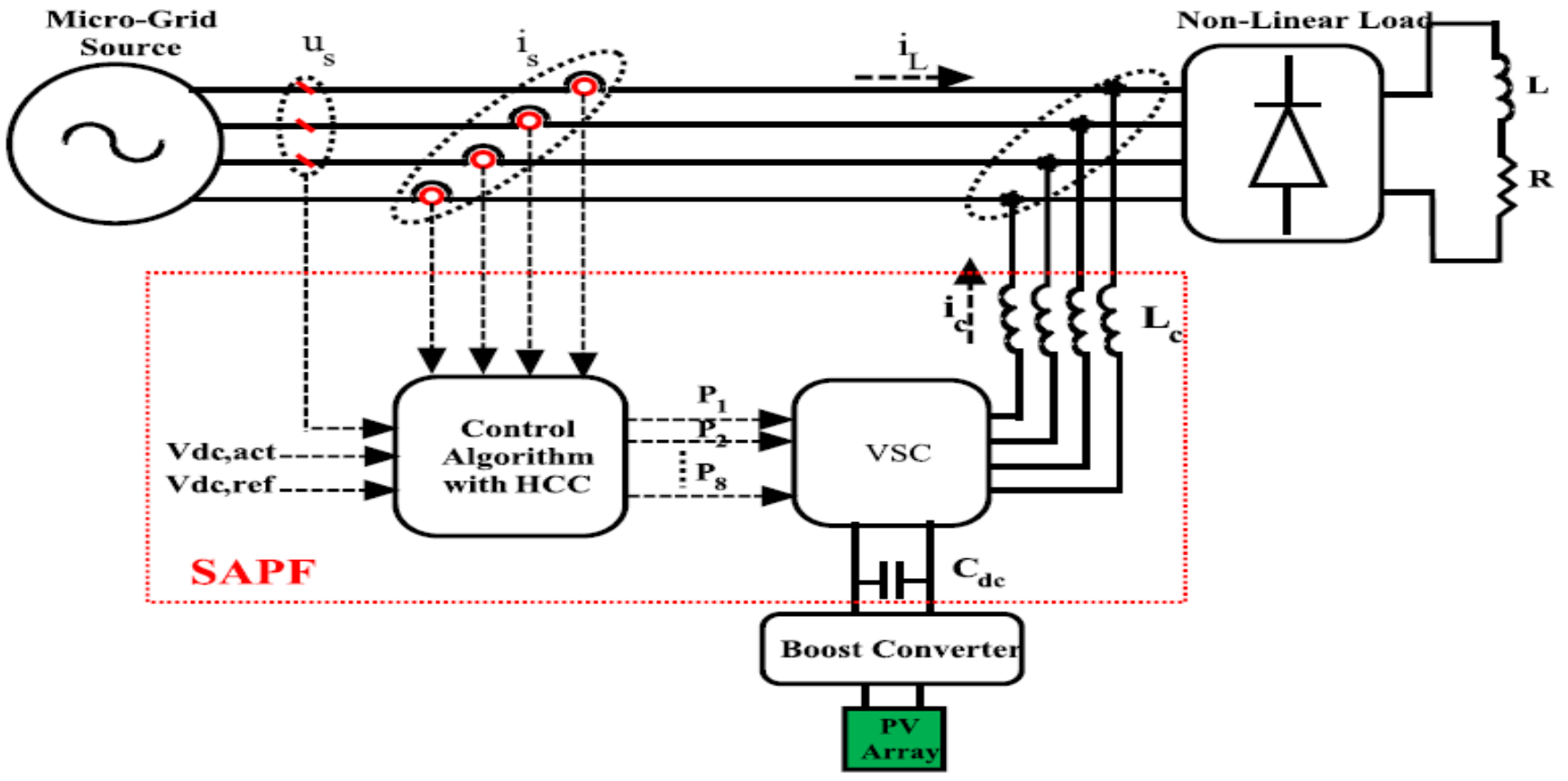

Fig.1 Configuration of SAPF interfaced with Micro Grid

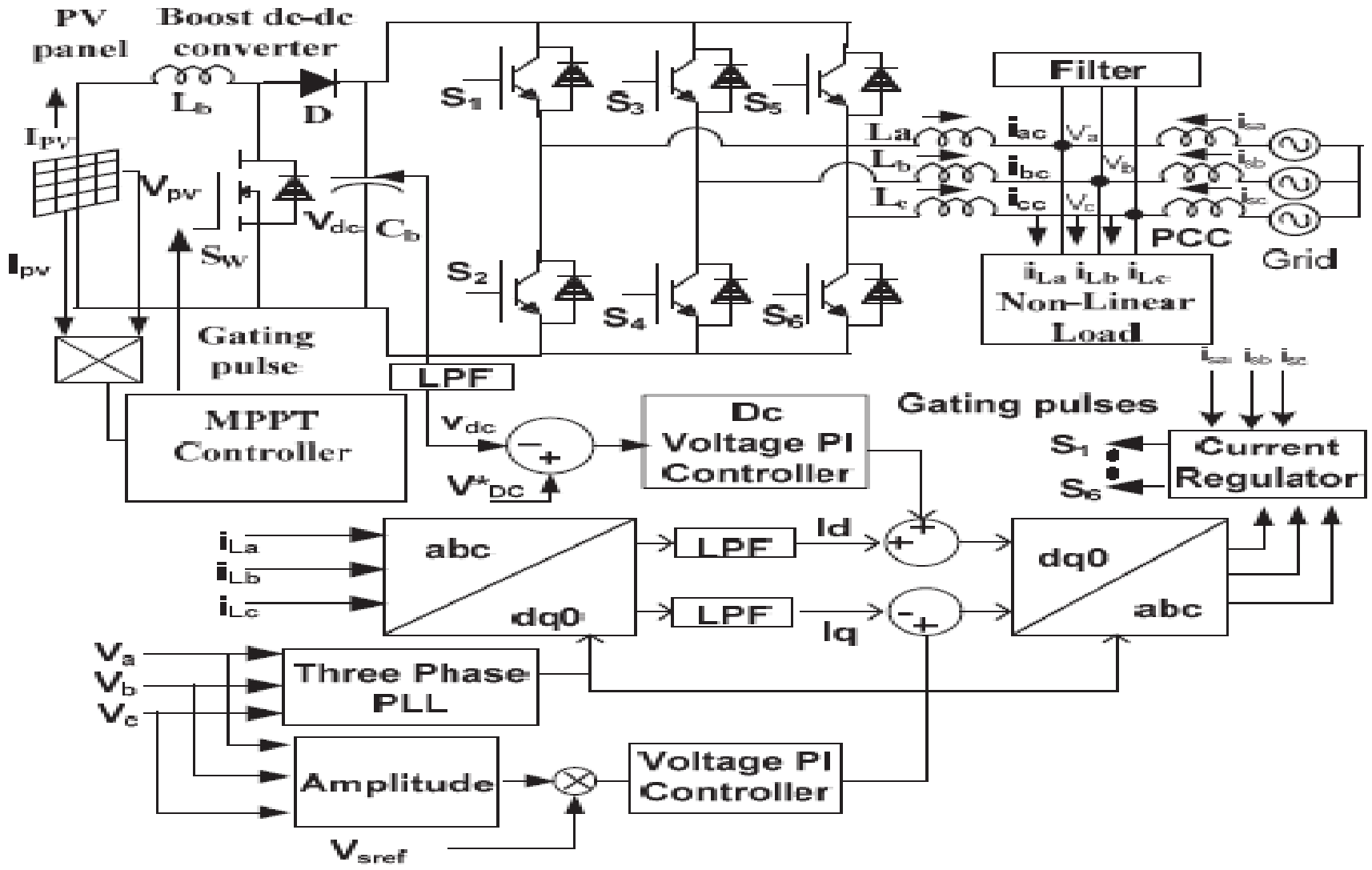

Fig.2 Schematic diagram of SPV interconnected with Grid System

The system consists of solar photo voltaic panel, DC-DC boost converter and voltage source converter. PV modules are arranged in series and parallel combination to get required voltage and current rating. By multiplying voltage and current characteristics, P-V characteristics are obtained and the point at which panel output is maximum (MPP) is tracked. By matching the load impedance MPPT is obtained. Therefore, a DC-DC converter is used to vary the input resistance of the panel by varying the duty cycle .In most of the shunt active power filter (SAPF), a voltage source converter is preferred over current source converters because of its high efficiency, low initial cost, compensates harmonic current, reactive power and balances the linear and nonlinear loads [7]. The System also employs PI controller to control the input voltage of the converter.

The Instantaneous current is given by

$$I_S = I_A + I_R + I_H \qquad (1)$$

Where $I_A$, $I_R$ and $I_H$ are active, reactive and harmonic load currents respectively.

The Instantaneous power is given by

$$P_S = P_A + P_R + P_H \qquad (2)$$

Where $P_A$ is active fundamental power , $P_R$ and $P_H$ are reactive & harmonic power flow in the system due to unbalanced and nonlinear load.

The active power drawn from the source is given by

$$P_A = P_s(t) * I_P * \sin wt \qquad (3)$$

Since there will be switching losses, conducting and leakage losses in a converter , these losses are supplied from Micro grid. The total peak current supplied from micro grid ($I_{SP}$) which includes loss current is given by

$I_{SP} = I_P + I_{Loss}$ (4)

There will not be any harmonics in the source current if the harmonic current and reactive current required by the load is supplied by shunt active Power filter. Total source current including losses is given by

$I_S^* = I_{SP} \sin wt$ (5)

Compensating current is given by

$I_C(t) = I_S - I_S^*$ (6)

Micro grid will supply necessary $I_S$* which is also known as the reference current. This current should be in phase with the supply line voltage. The three phase reference currents are given by

$I_{sa}{}^* = I_{SP} \sin\Theta$

$I_{sa}{}^* = I_{SP} \sin(\Theta - 120^o)$

$I_{sa}{}^* = I_{SP} \sin(\Theta - 240^o)$ (7)

## III. MPPT TECHNIQUE

The Solar panel converts 30-40% of incident light into electrical energy. To increase the efficiency of the Maximum Power point Tracking (MPPT) algorithm is implemented [8]. The Techniques like Perturb and Observe (hill climbing method) , Incremental Conductance method, Fractional short circuit current, Fractional open circuit voltage, Neural networks and Fuzzy logic are used to track the Maxim Power Point. Most commonly Perturb and observe (P&O)and Incremental conductance methods are preferred because of it is simple to implement and takes less time to track the Maximum Power point (MPP). MPPT algorithm calculates the reference voltage $V_{ref}$ towards which the PV operating voltage varied to the next value to get maximum power output. This process is repeated periodically at a rate of 1-10samples per second [9].

In this work Incremental conductance method is used and it is based on the principle that the slope of the PV array power curve is zero at the maximum power point. The MPP can be tracked by comparing the instantaneous conductance (I/V) to the incremental conductance ($\Delta I/\Delta V$). The whole system is simulated in MATLAB using Simulink. The source is taken as 3-phase 4-wire system and to get highly nonlinear characteristics, the nonlinear load is taken as rectifier with R-L load. Fig.3 shows the circuit of the Micro grid with nonlinear load [10]. In this work, an attempt has been made to study the dynamic performance of the system with an unbalanced load. The DC voltage of the Voltage source converter (VSC) has to be regulated and to be maintained constant under all loading disturbances. Since there is variation in the solar intensity and loading disturbances, the harmonics will be generated and appear in the source current. Therefore, the source current has to balanced and maintain sinusoidal in nature. This is achieved using an MPPT controller [10] and control approach for obtaining the harmonic free current waveform is discussed in the next section.

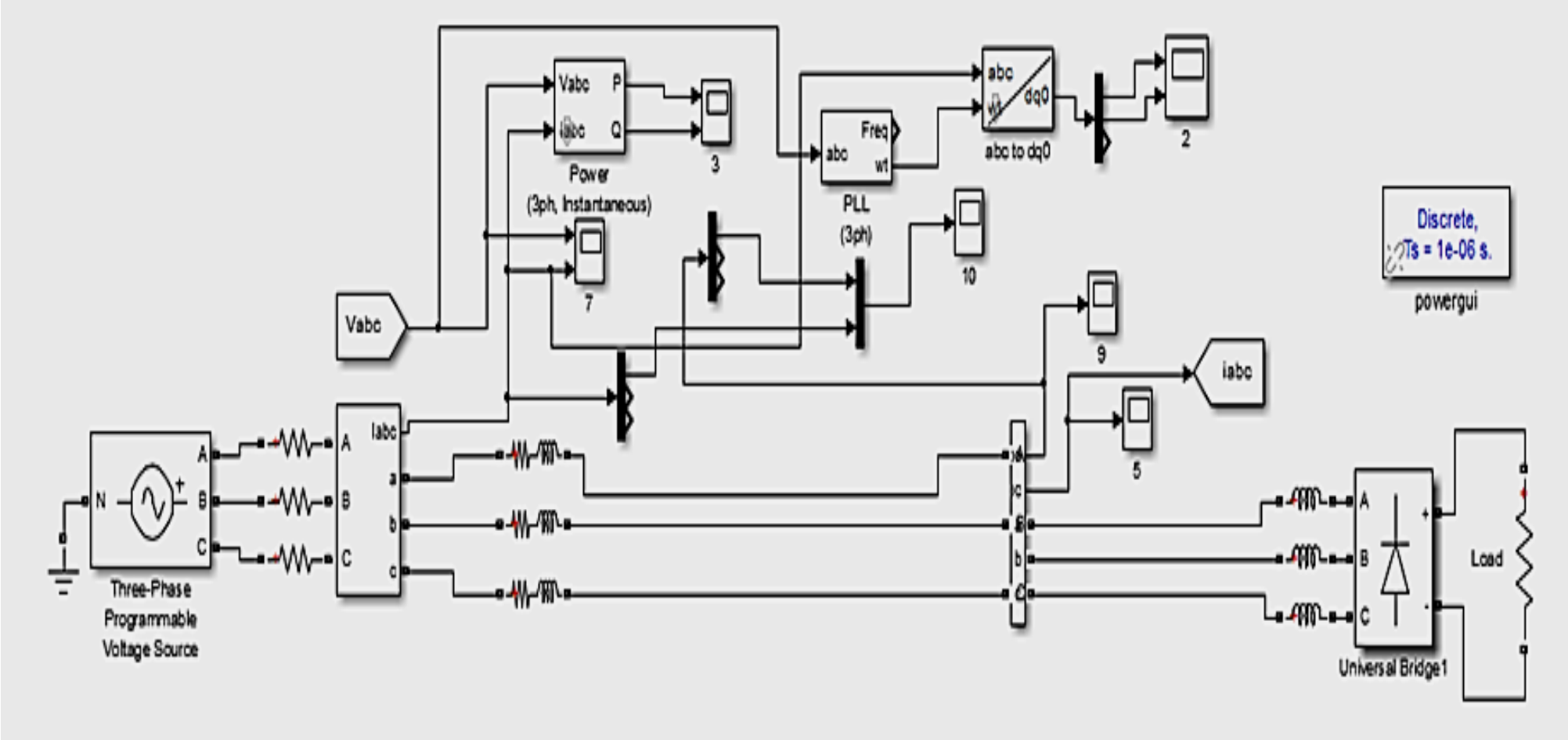

Fig.3Micro grid with nonlinear load

## IV. RESULTS AND DISCUSSIONS

Initially Simulation is performed without compensation, Source side voltage and current waveforms without compensation and Non- linear current waveform in the load side without compensation are captured and shown in fig.4 and fig.5 respectively. It has been observed that there is nonlinearity in the load side current. The compensator is switched on, the MPPT controller regulates/maintain the DC capacitor voltage at a given reference value. DC capacitor voltage will be passed through a low pass filter, which eliminates the ripple from the voltage waveform. This reference value governs the real power exchange between the SPV and Micro grid.

The compensating current generated by SAPF is shown in fig.6 and the Source side voltage& current waveforms with compensation is shown in fig.7. From fig.7, it is evident that the harmonics are eliminated in the load current waveform and the power quality is improved. FFT is performed to estimate the total harmonic distortion (THD).Fig.8 and Fig.9 shows the THD before and after compensation respectively.

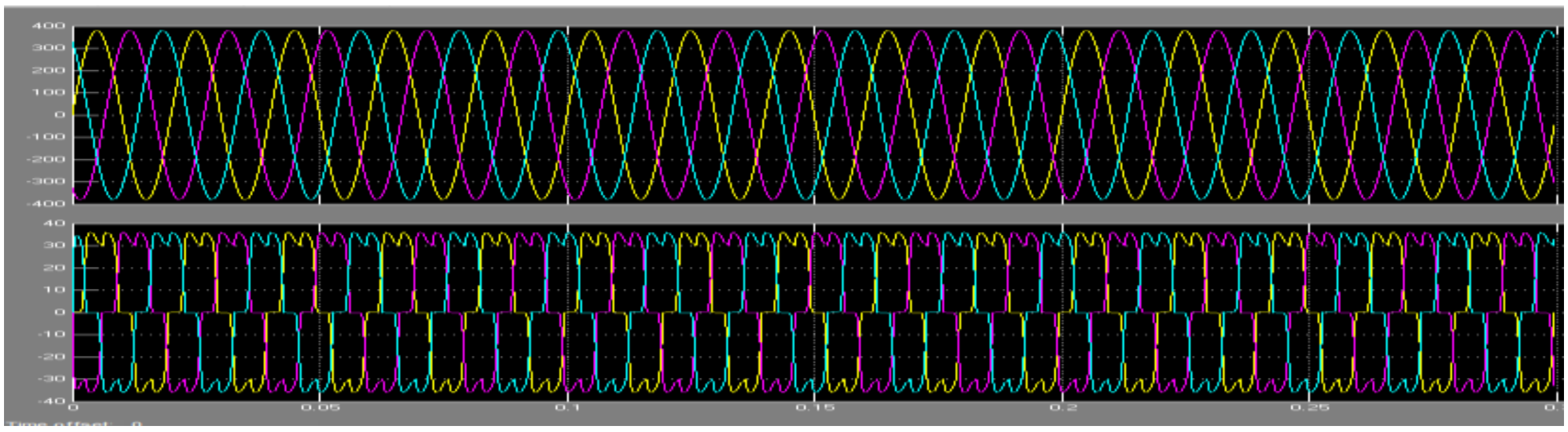

Fig.4Source side voltage and current waveforms without compensation

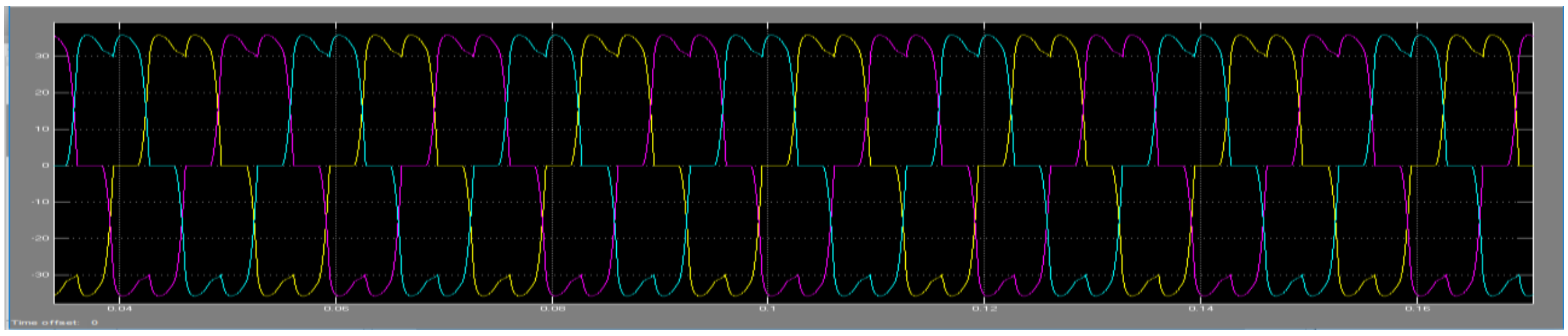

Fig.5Non linear current waveform in the load side without compensation

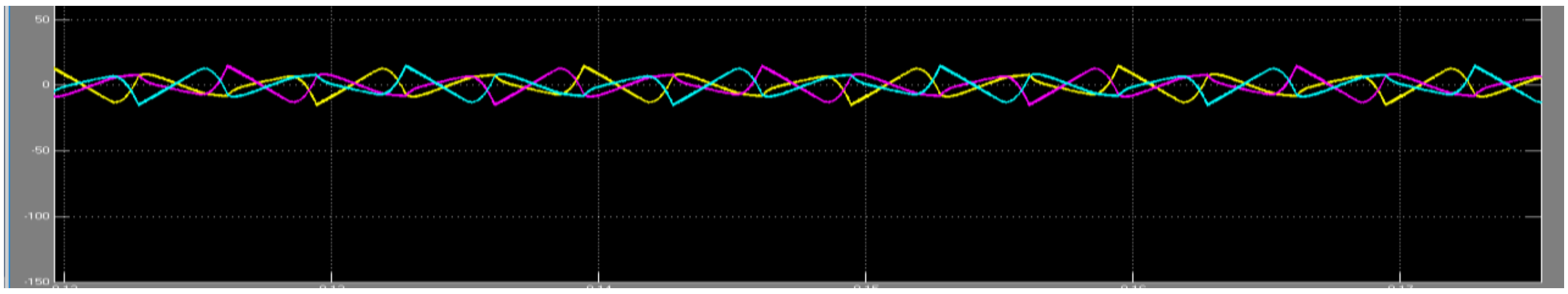

Fig.6 Compensating current generated by SAPF

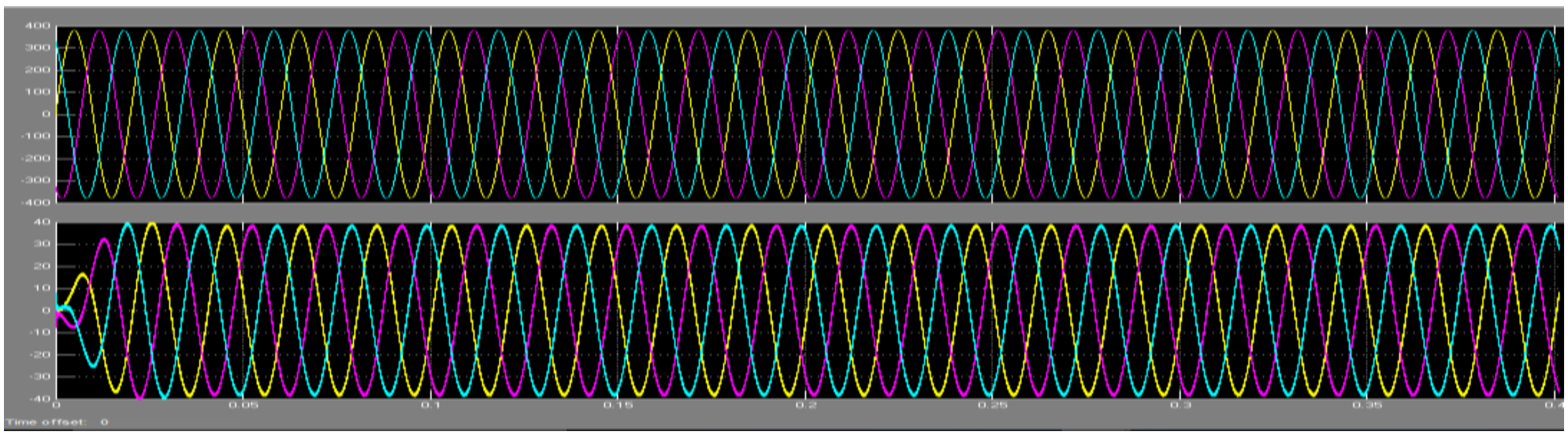

Fig.7Source side voltage and current waveforms with compensation

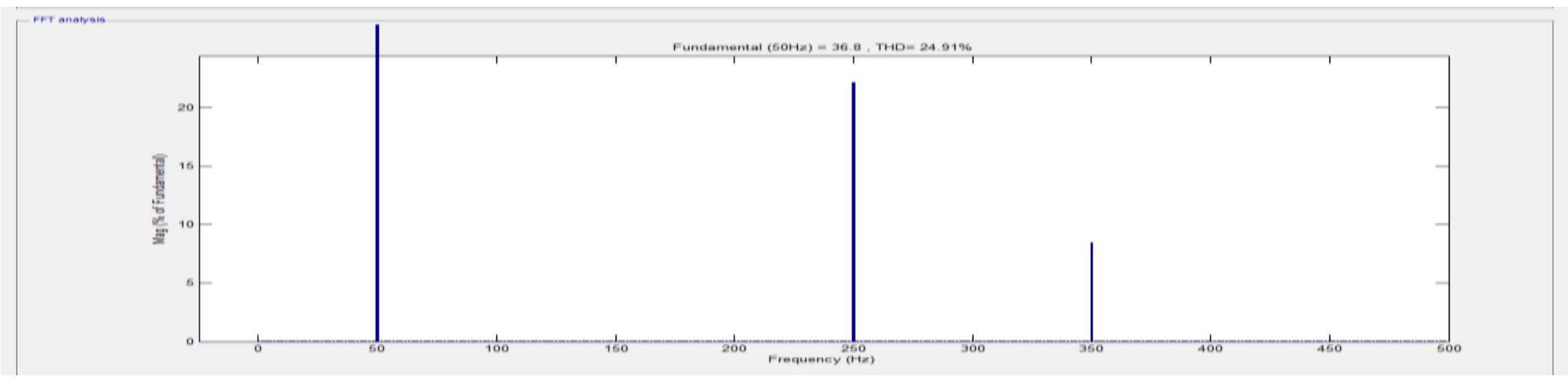

Fig.8Total Harmonic Distortion (THD) before compensation

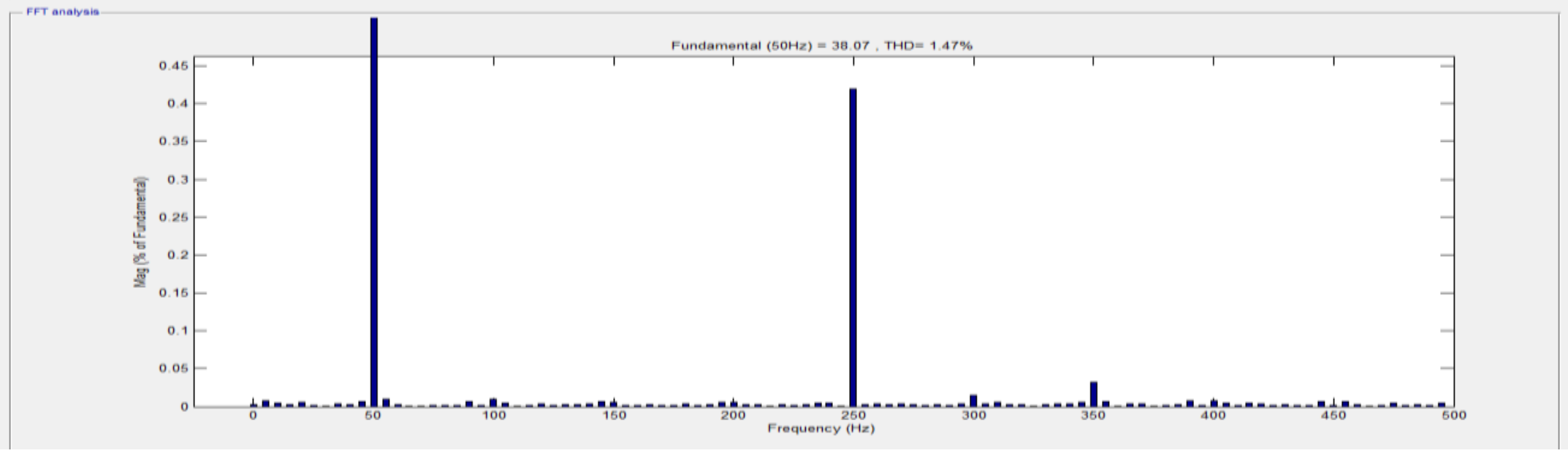

Fig.9Total Harmonic Distortion (THD) after compensation

## V. CONCLUSION

Modelling of Solar Photo voltaic array system interfaced with Micro grid using an MPPT controller is designed using the Simulink tool in MATLAB. The performance of the system with VSC at lagging power factor, non-linear load has been studied. Under different loading disturbances and varying solar intensity, harmonics generated in the system were eliminated by providing necessary compensation using an MPPT controller and Shunt Active Power filter combination. The load current obtained after compensation was free from harmonics and sinusoidal in nature, which is in phase with source voltage. There has been significant improvement in power quality of the system after compensation i.e. the THD was reduced from 25% to 1.5%.

## VI. ACKNOWLEDGEMENTS

The authors are thankful to the HOD, Department of Electrical & Electronics Engineering, Principal and Management of BMS Institute of Technology& Management, Bangalore for providing us necessary facilities, support and infrastructure to complete the work.